## Title

**Flow regime determines whether larval settlement hydrodynamics are governed by swimming speed or shape**

## Authors

Daniel Gysbers[1,†], Mark A. Levenstein[2], and Gabriel Juarez[3]

[1]*Department of Physics, University of Illinois Urbana-Champaign, Urbana, IL 61801, USA,*

[2]*Université Paris-Saclay, CEA, CNRS, NIMBE, LIONS, 91191, Gif-sur-Yvette, FR,*

[3]*Department of Mechanical Science and Engineering, University of Illinois Urbana-Champaign, Urbana, IL 61801, USA*

[†]Present address: Department of Physics, Adelphi University, Garden City, NY 11530, USA

Email: mark.levenstein@cea.fr

**ORCID**: DG, 0000-0001-7683-3728; MAL, 0000-0002-2309-3743; GJ, 0000-0002-5854-6925

**Abstract**

Larval dispersal and settlement govern the replenishment of sessile marine populations and the assembly of benthic communities, yet how larval traits interact with near-surface flow structures to direct settlement remains unclear. We used coupled hydrodynamic and larval-scale simulations to resolve how swimming speed and body shape, from circular to elongated, influence settlement in oscillatory flow over ridged substrates. Settlement success differed by up to an order of magnitude across trait combinations. Near-surface flow acted as a trait filter by differentially regulating transport to, retention near, and encounters with the substrate. The trait that mattered most depended on the local flow structure. Where rapid flow restricted access to settlement surfaces, swimming speed was most important. Where flow recirculated between substrate ridges, shape-dependent alignment controlled transport and retention. These results link larval traits to flow-mediated settlement mechanisms and provide a physical framework for understanding how recruitment varies across complex benthic habitats.

## Introduction

Many sessile marine invertebrates begin their lives as small motile larvae dispersed in the ocean that must eventually settle on a benthic substrate. These larvae, with normal body sizes ranging from hundreds of microns to a few millimeters, can be transported by currents over kilometers for up to several weeks before they reach a surface to attempt settlement (**Fig. 1a**) [1,2]. Due to these large spatial and temporal scales, it is difficult to directly observe this process in nature, and thus, gain an understanding of the mechanisms that promote and inhibit successful settlement [3,4]. In particular, before a larva can attempt to settle on a favorable substrate, it must first be physically transported to and access the benthic surface. This physical component of settlement occurs at small spatial scales near the substrate, where larval swimming interacts with the local flow field [5]. Knowledge of these physical mechanisms is therefore crucial for understanding settlement and recruitment, including choosing sites for ecosystem restoration and designing settlement or anti-fouling substrates.

Despite decades of research, the extent to which individual larval traits, such as shape and swimming speed, affect the physical component of settlement remains poorly understood. Much previous research has focused on the effects of (bio)chemical cues [6–12] or substrate roughness [13–18] on larval settlement preferences, with comparatively less attention given to how larval characteristics and local flow conditions interact to transport larvae to the substrate. Conversely, many studies investigating the effects of flow conditions on settlement have neglected the swimming ability of individual larvae, since larval swimming speeds are often much less than natural currents and wave action over coral reefs [19,20]. This motivates the common assumption that larvae can be treated as passive particles whose transport is determined primarily by external flow fields [5,21]. However, closer to benthic surfaces, eddies and velocity fluctuations can produce local flow speeds closer to larval swimming speeds, allowing larvae to actively influence their transport [19,22–25]. Previous studies have demonstrated that larvae and other zooplankton can respond to local flow conditions and chemical cues and actively control their position in the water column [23,26–30]. Thus, differences in larval swimming capability may become particularly important near the substrate.

The swimming speed of marine larvae can affect their ability to reach the substrate surface. Marine larvae exhibit a large range of swimming speeds, from coral species with average speeds of 1 to 5 mm $s^{-1}$ [19] to other benthic larvae, such as barnacles, reaching speeds of up to 2 cm $s^{-1}$ [23]. Even considering this wide range of swimming abilities, larval swimming speeds are still 1–3 orders of magnitude lower than typical currents over reefs [19]. In the same flow conditions, differences in swimming speed will therefore affect which flow regions larvae can access and the extent to which swimming contributes to transport toward the substrate.

Marine larvae also have a variety of sizes and shapes that affect their transport in flow. Using coral larvae as an example, larvae range from around 0.3 to 2.5 mm in length and have shapes from spherical to ellipsoidal or pear-shaped, with normal aspect ratios up to 5 (where aspect ratio is defined as length divided by width) [17,19,31,32]. Elongated swimmers in flows have been shown to align with the flow direction [33–35]. This alignment can affect transport by increasing trapping in vortices and preventing swimmers from crossing flow streamlines [33,35–38], such as elongated bacteria that can become trapped in high fluid shear regions [39]. Conversely, elongated plankton chains have been shown to have increased vertical transport in flow compared to lower aspect ratio single-cell counterparts [40]. The specific interactions between the shape of swimmers and flow structures in different environments can therefore affect transport to the benthic substrate and, in the case of marine larvae, the physical accessibility of settlement sites.

In previous work, our coupled flow-trajectory simulations established that, even for the same wave-driven oscillatory flow, substrate topography strongly affects larval settlement success [41]. As ridge height and spacing changed, near-surface flow shifted between skimming, detached, and reattached regimes (**Fig. 1b**), which changed whether larvae reached the surface by swimming alone or by a combination of swimming and vortices ejected by flow-topography interactions. Settlement success was the highest for millimeter-scale ridge geometries that produce flow near the transition between the detached and reattached regimes. However, this optimum was identified using larvae with a single prescribed swimming speed and shape. These flow regimes may therefore act as physical trait filters, favoring larvae whose swimming speed and shape allow them to reach and remain near the substrate. If so, then the topography that maximizes physical access to settlement surfaces could depend on larval traits as well as on the driving flow.

Here, our goal is to understand how the swimming speed and body shape of marine larvae influence their physical access to the substrate and subsequent probability of settlement. Using coupled hydrodynamic and larval-scale simulations to investigate a large experimentally inaccessible parameter space (**Fig 1c-e**), we demonstrate that near-surface flow acts as a hydrodynamic trait filter – altering settlement success by up to an order of magnitude depending on the specific trait combination. Analysis of local flow speeds and larval concentration profiles reveals the interaction between larval swimming and high flow transport barriers, which inhibit larval movement to the substrate surface and their subsequent settlement. Settlement heatmaps confirm the effects of swimming and flow interactions near the surface and illustrate the impact of lower flow speed regions near ridge walls. An analysis of larval alignment with the flow demonstrates the effect of larval aspect ratio on transport rates to the surface. Finally, concentration and settlement rate analyses explain the reduced influence of swimming speed on settlement for some substrate topographies. These results provide evidence that larval traits can alter the physical accessibility of settlement habitats for otherwise identical flow and substrate conditions and may help explain differences in natural recruitment patterns and inform more targeted interventions for ecosystem restoration.

**Methods**

***Modelling flow.*** A finite-element model was developed in COMSOL Multiphysics to simulate fluid flow above substrates with different rectangular ridged topographies (**Fig. 1c**) [24,41]. Substrate topographies had a ridge height, $h$, of 3 mm and a spacing between ridges, $w$, from 0.75 to 90 mm to investigate $w/h$ values of 0.25, 0.5, 1, 2, 4, 10, 20, and 30. The 2D geometries were created in COMSOL with a no-slip bottom wall, top symmetry condition, and periodic side boundaries to simulate an infinite length of repeating ridges. The height of the simulated region was varied between 20 and 60 mm to avoid edge effects from ejected vortices, with 20 mm being used unless a taller domain was required. The simplified, purely oscillatory flow was driven by a sinusoidal pressure difference on the periodic side boundaries with a period ($\tau$) of 5.5 s and an amplitude dependent on the simulated area to yield a max flow speed $U_{max} \approx 60$ mm $s^{-1}$, chosen as a representative reef flow condition [41]. While the applied max flow was 60 mm $s^{-1}$, local flow velocities can exceed this value due to the flow constriction caused by the presence of ridges. The Reynolds number for these simulations is between 1000 and 4000, calculated from the channel half-height; therefore, the flow is nominally laminar. As validated previously, we used an average mesh density of 159.36 elements $mm^{-2}$ and a time step of 0.01 s for the flow simulation [24]. The flow fields produced after 50 periods were utilized to avoid startup effects, and 10 consecutive periods were imported into MATLAB to simulate individual larval trajectories. More details on the flow model can be found in ref. [24] and ref. [41].

The applied oscillatory flow cycles through a repeating sequence of phases: forward acceleration, deceleration, reversal, backward acceleration, deceleration, and a second reversal that begins the next period. Because the forward and backward halves of the cycle are symmetrical, we focus our analysis on two characteristic phases: "max flow" when the flow speed peaks in either direction (at $t = \tau/4$ and $3\tau/4$), and the "turning point" when the flow reverses direction (at $t = 0$, $\tau/2$, and $\tau$).

The substrate topographies chosen produced three distinct flow regimes, distinguished by the behavior of the recirculating vortical structure within the ridge gap during the max and turning point flow phases [41] **(Fig. S1)**. We refer to these as skimming flow (regime I), detached flow (regime II), and reattached flow (regime III). In skimming flow (regime I), a vortical structure fills the gap during max flow and persists through the turning point, limiting flow into and out of the gap. In detached flow (regime II), a vortical structure similarly fills the gap during max flow but breaks down and is ejected at the turning point, allowing streamlines to cross between the gap and flow above the ridges. Finally, in reattached flow (regime III), a vortical structure forms at the upstream ridge during max flow but does not fill the entire gap, allowing the flow to reattach to the substrate. These three flow regimes, and the representative *w*/*h* substrate in each regime (0.5, 4, and 10), form the basis of the flow and settlement analyses presented in the Results section.

***Simulating larval transport and settlement.*** Larval trajectories were simulated with a custom 2D individual-based model developed in MATLAB [24,41]. Larvae were modeled as neutrally buoyant ellipses with a constant swimming velocity, $u_l$, along their major axis (**Fig. 1d**). Swimming speeds of $u_l$ = 1, 2, 3, and 4 mm $s^{-1}$ were chosen to span the range of swimming speeds measured in different coral species [19]. For model simplicity, and to avoid making assumptions about larval behavior, we do not consider the ability of larva to modify their speed, buoyancy, or orientation in response to (bio)chemical or hydrodynamic cues. Thus, our simulated larvae—while motile—are not responsive to their environment. The larvae were simulated as straight swimmers that were started in a grid of 30 positions (at heights of 8, 10, and 12 mm from the surface at 10 equally spaced x-positions above the substrate; $x_i$, $y_i$), 32 initial orientations ($\theta_i$), and 20 initial times equally spaced throughout the first 5.5 s flow period ($\tau_i$, **Fig. 1c, Table S1**). Translational motion of the larvae was calculated using the equation:

$$\dot{\boldsymbol{r}} = \boldsymbol{u} + u_l\hat{\boldsymbol{n}},$$

where $\dot{\boldsymbol{r}}$ is the total larval velocity, $\boldsymbol{u}$ is the local flow velocity, and $\hat{\boldsymbol{n}}$ is the unit vector along the larval major axis. Larval rotation was calculated using the equation:

$$\dot{\theta} = \frac{\omega_z}{2} + \alpha\hat{\boldsymbol{g}} \cdot \boldsymbol{E}\hat{\boldsymbol{n}},$$

where $\dot{\theta}$ is the total angular velocity, $\omega_z = \hat{\boldsymbol{z}} \cdot (\nabla \times \boldsymbol{u})$ is the vorticity, $\alpha$ is a shape parameter, $\boldsymbol{E} = (\nabla\boldsymbol{u} + \nabla\boldsymbol{u}^T)/2$ is the symmetric rate of strain tensor, and $\hat{\boldsymbol{g}}$ is the unit vector along the larval minor axis. Larvae were represented as two-dimensional ellipses with semimajor axis *a*, semiminor axis *b*, and aspect ratio $AR = a/b$ **(Fig. 1d)**. To account for the effect of shape on a larva's rotational response to the local rate of strain, we used the shape parameter, $\alpha = (AR^2 - 1) / (AR^2 + 1)$. For the elliptical shapes considered here, $\alpha$ is 0 for a circular larva and approaches 1 as the larva becomes increasingly elongated. Therefore, elongated larvae exhibit a stronger rotational response to the local strain field than circular larvae. We simulate larvae with $a$ = 0.5 mm and $b$ = 0.5, 0.388, 0.288, 0.188, and 0.114 mm, which resulted in an $\alpha$ = 0, 0.25, 0.5, 0.75, and 0.9, respectively (**Fig. 1e)**. These values correspond to aspect ratios of $AR$ = 1, 1.29, 1.74, 2.66, and 4.39, respectively, and represent realistic larval shapes (**Table S2**) [19,24,30–32,42–45].

Larvae started with an initial position, orientation, and phase in the flow period, and their subsequent motion was governed by the translation and rotation equations until they "settled" on a surface, exited from the top of the simulated domain, or more than 10 periods had passed without settlement. Larvae were considered settled and removed from the simulation when they encountered a surface while their total speed ($\dot{\boldsymbol{r}}$) was lower than 1.33× the larval swimming speed to prevent settlement at unrealistically high speeds. The 1.33×$u_l$ settlement cutoff was chosen based on the species-averaged coral larval swimming speed of 3 ± 1 mm $s^{-1}$ (mean ± standard deviation) from ref. [24] and kept proportional for the other investigated swimming speeds. Larvae remained in the water column if they encountered a surface with a speed larger than the cutoff. We neglect post-attachment effects, and an implicit assumption of our model is that the simulated larvae desire to and are competent to settle on surfaces they encounter, as long as the speed cutoff is met. The rationale behind this decision was to isolate hydrodynamic factors from biological or chemical factors and focus exclusively on the end of the pelagic phase of the coral life cycle. Additionally, each larva was simulated independently, and interactions between larvae were not considered.

Since larvae were injected into the flow only during the first flow period and then left to swim and settle, the number of larvae being transported in the flow quickly decreased as larvae settled or exited out the top of the simulation area, affecting transport rate analysis. To investigate transport and settlement rates more completely, additional simulations were conducted with larvae constantly injected into the flow across the full 10 flow periods. These larvae were initiated in a grid of 15 starting positions (at heights of 8, 10, and 12 mm from the surface at 5 equally spaced x-positions above the substrate; $x_i$, $y_i$), 32 initial orientations ($\theta_i$), and 100 initial times equally spaced throughout 55 s (10 periods). Constantly injected larvae were simulated until they settled on a surface, exited from the top of the simulated domain, or the 55 s time cut-off from the first set of larvae initialized was reached.

***Statistical Analysis.*** The simulated settlement results were analyzed at the condition-level using a beta regression model with a logit-linked mean to understand the importance of different variables in each flow regime and predict settlement success across the parameter range [46]. Several models of increasing complexity were evaluated and the final fitted model given below was selected using the "one standard error" rule [47]:

$$\text{logit}(\mu) = C_0(R) + C_{w/h}(R)X + C_s(R)S + C_\alpha(R)A + \beta_{s\alpha}SA + I_{R=3}(\gamma_1 X^2 + \gamma_2 XA + \gamma_3 X^2 A)$$

where $\mu$ is the predicted success proportion, $C$ values are fitted coefficients that vary as a function of flow regime ($R$), $X$ is the natural log of standardized *w*/*h*, *s* is the swimming speed, $S$ and $A$ are the standardized speed and $\alpha$, respectively, and $\beta_{s\alpha}$ is the fitted coefficient for speed:$\alpha$ interactions. $I_{R=3}$ is an indicator function that is equal to 0 for flow regimes I and II and equal to 1 for flow regime III, which allows fitting the non-linear behavior observed in regime III with a quadratic function. The γ values are fitted coefficients for the regime III quadratic function. Models were evaluated based on their out-of-sample predictive performance **(Fig. S2)**, and the relative importance of simulated parameters (*w*/*h*, speed, $\alpha$, and speed:$\alpha$ interactions) was assessed through the increase in cross-validated root mean squared error (CV RSME) upon parameter removal. We additionally employed principal components analysis (PCA) as an unsupervised model-independent method to compare with results from the beta regression. A more detailed description of the statistical analysis and the Python script utilized are included in the Supplementary Materials [48–54].

## Results

***Larval settlement trends and statistical analysis.*** The modelled larval characteristics influence settlement success—defined as the percent of larvae that settled of the total number of larvae simulated—for a given substrate topography (*w*/*h*) under the simulated oscillatory flow condition ($U_{max}$ = 60 mm $s^{-1}$, $\tau$ = 5.5 s) **(Fig. 2a)**. For the different topographies, swimming speeds, and shapes simulated in this study, the larval settlement percentages range from a minimum of 0.7% to a maximum of 52.7%. Since the set of substrate topographies utilized produces three distinct flow regimes—skimming flow, detached flow, and reattached flow **(Fig. S1)** [41]—it allows us to examine the effects of the same physical larval characteristics under different hydrodynamic conditions. For clarity, the figures and analyses in the main text are focused on larvae with speed $u_l$ = 1 mm $s^{-1}$ and $u_l$ = 4 mm $s^{-1}$ and a representative *w*/*h* in each flow regime: 0.5, 4, and 10. The full settlement results are available in the Supplementary Information **(Table S3 and Fig. S3a)**. The entire dataset was fit with a beta regression model to predict settlement success across the investigated parameter space (**Fig. S3b and S4**) and evaluate the predictive performance of the input parameters, *w*/*h*, speed, $\alpha$, and speed:$\alpha$ interactions, in each flow regime (**Fig. 2b**).

In the skimming flow regime, larval settlement is the lowest overall, ranging from 0.7% to 14.3%, and success does not vary with the *w*/*h* value of the substrate until approaching the transition to detached flow at *w/h* = 1 (**Fig. 2a, I**). Settlement percentage also did not appear to be affected by larval shape but was clearly influenced by their swimming speed. For example, settlement on the *w/h* = 0.5 substrate for larvae with $u_l$ = 1 mm $s^{-1}$ ranges between only 0.7% for $\alpha$ = 0 and 1.2% for $\alpha$ = 0.9 yet increases by an order of magnitude for larvae with $u_l$ = 4 mm $s^{-1}$, ranging between 9.4% for $\alpha$ = 0 and 12.3% for $\alpha$ = 0.9. These trends are supported by the beta regression model, which shows little predictive importance of *w*/*h* and $\alpha$ and more importance of speed in this flow regime (**Fig. 2b**)

In the detached flow regime, settlement success increases for all larval parameter combinations compared to under skimming flow, ranging from 7.6% to 34.9% **(Fig. 2a, II)**. Settlement also increases with increasing *w/h* for all larval types, and *w*/*h* is shown to have more predictive importance (**Fig. 2b**). Larvae with the same swimming speed have large variations in settlement success in this regime, indicating settlement is not dominated by speed. More specifically, settlement success on the representative *w/h* = 4 substrate for larvae with $u_l$ = 1 mm $s^{-1}$ and $u_l$ = 4 mm $s^{-1}$ has almost the same range of 17.8% to 31.3% and 14.1% to 27.3%, respectively, with the larger $\alpha$ larvae having increased settlement. In particular, larvae with the same $\alpha$ have similar settlement percentages regardless of their swimming speed, indicating that settlement is more sensitive to shape than speed in this regime as supported by the beta regression model (**Fig. 2b**).

Larval settlement success reaches a maximum in the reattached flow regime, with percentages ranging from 12.3% to 52.7% **(Fig. 2a, III)**. While the overall maximum settlement percentage increases in this regime, larvae with $u_l$ = 1 mm $s^{-1}$ have lower success than in detached flow with a maximum of only 23.5%. Faster swimming larvae reach their maximum settlement percentage at either *w/h* = 10 or *w/h* = 20, depending on their shape, and then experience decreasing settlement with increasing *w/h* consistent with previous results [41]. Like in the skimming flow regime, larvae with the same swimming speed have similar settlement success, demonstrating that settlement hydrodynamics are also dominated by swimming speed under reattached flow. However, there is still variation due to shape. For example, at *w/h* = 10, settlement percentages for larvae with $u_l$ = 1 mm $s^{-1}$ and $u_l$ = 4 mm $s^{-1}$ range from 19.8% for $\alpha$ = 0 to 23.5% for $\alpha$ = 0.9 and from 34.6% for $\alpha$ = 0 to 52.7% for $\alpha$ = 0.9, respectively. These trends are also reflected in the beta regression model, where *w*/*h* and speed have strong predictive importance and $\alpha$ is more important to model accuracy than in skimming flow (**Fig. 2b**). Interestingly, speed:$\alpha$ interactions have no predictive importance in any of the three flow regimes, implying that speed and shape influence settlement hydrodynamics independently. These results are in agreement with an independent principal components analysis (PCA) performed on complete settlement success curves as a function of *w*/*h* (**Fig. S5**).

The above results demonstrate that settlement success varies with physical larval traits, with larval speed and shape having different effects according to the particular flow regime. To better understand the mechanisms behind these settlement trends, we explore the unique impact of speed and shape on larva-scale hydrodynamic interactions and larval transport in each flow regime. Beyond the aggregate settlement success values, the individual-based simulation gives us access to complete larval trajectories, which are used to analyze and compare transport across the simulated parameter space. In the following sections, we assess changes in larval concentration profiles, surface availability, and settler density (skimming and reattached flows) alongside flow alignment, transport and settlement rates (detached flow) to explain how this hydrodynamic trait filter governs physical access to settlement refugia across distinct flow environments.

***Skimming flow: Transport barriers limit settlement.*** For a substrate with $w/h = 0.5$, representative of the skimming flow regime, settlement is dominated by larval swimming speed **(Fig. 2a, I and 2b)**. In this regime, the flow above the ridge tops is fast throughout the majority of the period. For example, speeds reach 63 mm s$^{-1}$ during max flow and reach 19 mm s$^{-1}$ during the turning point flow **(Fig. 3a)**. Within the sheltered gaps between ridges, flow speeds are much lower, never surpassing 1.2 mm s$^{-1}$. However, flow streamlines in the bulk are isolated from the vortices in the gaps, and the high flow speed region above the ridges creates a transport barrier that larvae must cross to enter the gaps and settle **(Fig. S1a)**.

The effect of this transport barrier can be seen in the concentration profiles of larvae in the water column **(Fig. 3b** and **3c)**. Larvae of all investigated characteristics have a concentration peak just above the ridge tops where they are stopped by the transport barrier. This effect is strongly dependent on larval swimming speed, with the concentration of slower larvae increasing as larvae are unable to move beyond the barrier. For example, after 3 flow periods, larvae with $\alpha = 0.5$ and 1 mm s$^{-1}$ swimming speed have a concentration peak of 2.67% of the total larvae situated 2 mm above the ridge tops while 4 mm s$^{-1}$ swimming larvae have a concentration peak of just 0.08% at 1 mm above the ridges, an over 33× decrease **(Fig. 3b)**. This effect is less dependent on larval shape, with concentration peaks for larvae with the same 3 mm s$^{-1}$ swimming speed ranging from 0.46% for larvae with $\alpha = 0$ up to 1.11% for larvae with $\alpha = 0.9$ **(Fig. 3c)**. Similarly, the location of this peak remains unchanged at 0.5 mm above the ridge tops for all larval shapes, implying the barrier height is also determined by the swimming speed.

The high flow speeds above the ridges impede larvae from entering the sheltered gaps to settle, causing many larvae to either settle on the ridge tops or concentrate above the ridges without settling. This is illustrated by exploring the particular locations where the larvae are able to settle **(Fig. 3d)**. For the example of larvae with $\alpha = 0.5$, slow larvae with $u_l = 1$ mm s$^{-1}$ have very little settlement anywhere on the substrate, with a maximum settlement density of 40 larvae mm$^{-1}$ on the top of the ridges and 22 larvae mm$^{-1}$ within the gaps. As larval speed increases, more larvae are able to settle on the tops of the ridges during the short time windows when the periodic flow reverses and flow speed decreases. Larvae with $u_l = 2$ mm s$^{-1}$ have a maximum settler density of 312 larvae mm$^{-1}$ on the tops of the ridges and larvae with $u_l$ = 4 mm s$^{-1}$ have a maximum density of 680 larvae mm$^{-1}$. Faster swimming larvae are also better able to cross the high flow speed transport barrier above the ridges to settle inside the gaps. Larvae with $u_l$ = 2 mm s$^{-1}$ have a maximum settler density of 218 larvae mm$^{-1}$ inside the gap, increasing to 256 larvae mm$^{-1}$ for larvae with $u_l$ = 3 mm s$^{-1}$ and to 362 larvae mm$^{-1}$ for larvae with $u_l$ = 4 mm s$^{-1}$. However, settlement success is still quite low, even for faster larvae, compared to in the other flow regimes **(Fig 2a)**.

***Reattached flow: Near-surface flow limits settlement.*** For a substrate with *w*/*h* = 10, representative of the reattached flow regime, settlement is strongly influenced by larval swimming speed, while larval shape also plays a role **(Fig. 2a, III and 2b)**. Wider gaps on these substrates are less sheltered by the ridges, meaning that while streamlines can enter into the gaps to increase transport in and out, flow speeds remain high throughout the flow period **(Fig. S1c)**. The flow inside the gaps reaches a maximum speed of 27 mm s$^{-1}$ during both the max flow and turning point phases of the flow (**Fig. 4a**). Even at only one larval length from the surface (1 mm in our case), flow speeds can be as high as 11 mm s$^{-1}$. This high flow speed causes the area in the center of the gap to act as an unsheltered, flat substrate, impeding settlement.

Larval concentration profiles are examined to investigate the transport or settlement barriers encountered by the larvae **(Fig. 4b, 4c)**. All larvae develop a concentration peak just above the bottom surface of the gaps, and this peak decreases at higher larval swimming speeds. For example, after three flow periods, larvae with $\alpha$ = 0.5 and swimming speed of 1 mm s$^{-1}$ have a near-surface concentration peak of 2.18% while larvae with the same shape and a speed of 4 mm s$^{-1}$ have a concentration peak that is more than a factor of two lower (0.93%) **(Fig. 4b)**. The near-surface concentration also varies with larval shape **(Fig. 4c)**. At *w*/*h* = 10, larvae with larger α have higher concentration peaks near the surface, with peak concentrations increasing from 0.74% for α = 0 to 1.62% for α = 0.9. However, unlike the relationship observed with swimming speed, this increase in near-surface concentration does not correspond to a decrease in settlement. Instead, the settlement percentages at *w*/*h* = 10 increase with α **(Fig. 2a, III)**. Larval shape therefore has a more complex effect in the reattached flow regime. At *w*/*h* = 20, the effect of $\alpha$ on settlement reverses, indicating that the influence of larval shape in the reattached regime depends on substrate topography (**Fig. 2b and S5b**).

The near-surface accumulation reflects larvae being transported into the gaps but unable to settle during much of the flow period because of high local flow speeds. Larvae that do not settle can remain within the gaps, contributing to the observed concentration peaks **(Fig. 4c)**. We set a settlement speed threshold of 1.33× the larval swimming speed to prevent unphysical settlement (see **Methods**), so flow speeds in the gaps of over 5× the swimming speed of even the fastest swimming larvae prevent settlement except during lulls in the flow. To understand and quantify this effect, we computed the percentage of the 1D surface length within the gaps that would be available to larvae for settlement based on the flow speed over a full period (**Fig. 4d**). For slow swimming, $u_l$ = 1 mm s$^{-1}$ larvae, the maximum surface length available for settlement is 15.33%, and less than 5% is available for most of the period. Fast swimming larvae with $u_l$ = 4 mm s$^{-1}$ have a larger percentage of substrate available, with a maximum of 24.67%.

The effects of high flow speeds in the gaps are further seen in the settlement locations of the larvae (**Fig. 4e**). For representative larvae with α = 0.5, slow larvae with $u_l$ = 1 mm s$^{-1}$ have a maximum settlement density of 222 larvae mm$^{-1}$ in the more sheltered corners of the gap (the 25% of length at each side) and only 89 larvae mm$^{-1}$ in the exposed center 50%. Fast larvae with $u_l$ = 4 mm s$^{-1}$, have the highest maximum settlement density of 337 larvae mm$^{-1}$ in the corners and 216 larvae mm$^{-1}$ in the exposed center.

***Detached flow: Shape-dependent transport governs settlement.*** For *w*/*h* = 4, representative of the detached flow regime, settlement success is more strongly influenced by larval shape than by larval speed, unlike the other two flow regimes **(Fig. 2a, II and 2b)**. In detached flow, similar to reattached flow, vortex ejection at the turning points allows fluid streamlines to cross between the bulk flow and gaps. However, like skimming flow, a slow-moving vortex fills the gaps at max flow, with flow speeds from 4 to 8 mm s$^{-1}$, sheltering them from the flow above the ridges **(Fig. S1b).**

Larval shape affects near-surface transport by influencing the alignment of larvae with the flow (**Fig. 5a**). Above *w*/*h* = 4 substrates, increasing $\alpha$ increases the amount of time larvae spend aligned with the flow, with larvae of $\alpha$ = 0.9 spending a larger percentage of time aligned with the flow than spherical larvae ($\alpha$ = 0) at all swimming speeds. Within 5 mm of the bottom surface, larvae with $\alpha$ = 0 align with the flow for 10.5% to 12.5% of the time, while larvae with $\alpha$ = 0.9 align for 17.5% to 23.1% of the time for faster to slower swimming larvae, respectively. This difference in alignment affects how the larvae of different shapes interact with the vortices and flow over the ridges, influencing their near-surface transport. These interactions can be visualized by overlaying all larval trajectories, which clearly shows that elongated larvae more closely follow recirculatory streamlines within and above ridge gaps (**Fig. S6** and **Video S1**).

To quantify this effect directly, we calculate the net rate of larval transport into the ridge gaps. Simulations with constant, regular larval injections are used for this analysis because the number of larvae in the settlement simulations decreases as larvae settle or exit out the top of the simulation area, which affects the transport rate. The net transport rate is calculated as the difference between the rates of larvae entering and exiting the gap, normalized by the total rate of larvae entering the simulation. Net transport rates into the gaps range from 15.5% to 23.3% for larvae with $\alpha = 0$, whereas larvae with $\alpha = 0.9$ have the highest transport rates, ranging from 27.5% to 35.7% (**Fig. 5b**). The increased transport into the gaps with increasing $\alpha$ is correlated with increased settlement (Pearson's $r = 0.95$, $p < 0.001$). In general, larvae with characteristics that produce higher net transport rates have higher overall settlement percentages **(Fig. 5b)** Thus, in detached flow, one way larval shape affects settlement is through its influence on near-surface transport into the ridge gaps.

Swimming speed produces a different effect in this regime. Although swimming speed has little effect on the overall settlement percentages for larvae with the same α **(Fig. 2a, II)**, it affects both the number of larvae retained near the surface and their ability to settle once there. For constantly injected larvae, slower swimming larvae accumulate at higher densities within the ridge gaps than faster larvae **(Fig. 5c, left)**. Across all α values, larvae with $u_l = 1$ mm $s^{-1}$ reach average densities of 61 to 81 larvae $mm^{-1}$ in the gap region, compared with only 7 to 10 larvae $mm^{-1}$ for larvae with $u_l = 4$ mm $s^{-1}$. During the turning points of the flow period, vortices are ejected from the ridge gaps and transport larvae between the bulk flow and gap region. During max flow, higher flow speeds near the ridge tops act as a transport barrier that traps larvae in the gaps. This trapping is more effective for slower larvae, resulting in greater accumulation near the surface.

However, the greater near-surface density of slower larvae does not necessarily result in greater settlement. The fractional settlement rate, measured relative to the number of larvae in the gaps, increases with swimming speed **(Fig. 5c, right)**. Larvae with $u_l = 1$ mm $s^{-1}$ have fractional settlement rates of 0.05 to 0.07 $s^{-1}$, whereas larvae with $u_l = 4$ mm $s^{-1}$ have rates of 0.35 to 0.45 $s\text{-}^{1}$, more than 6× higher. Thus, faster larvae are less frequently retained in the gaps but are more likely to settle while they are there. The higher flow speeds within the gaps compared with the skimming flow regime inhibit settlement of slower larvae.

These opposing effects partially compensate for one another when the absolute settling rate is considered **(Fig. 5d)**. The greater near-surface density of slower larvae is offset by their lower fractional settlement rate, while faster larvae have lower near-surface density but higher fractional settlement rates. However, these effects do not completely cancel. Instead, the combination produces a maximum in absolute settling rate at an intermediate swimming speed, with larvae at $u_l = 2$ mm $s^{-1}$ generally exhibiting the highest settlement rates for $\alpha \geq 0.5$ **(Fig. 5d)**. For example, larvae with $\alpha = 0$ have settling rates from 135 to 178 larvae $s^{-1}$ across the investigated swimming speeds, while larvae with $\alpha = 0.9$ have settling rates from 241 to 310 larvae $s^{-1}$. Thus, swimming speed does influence settlement in the detached flow regime. However, its effect is smaller than that of larval shape, which in addition to increasing transport, increases both larval retention and the fractional settling rate (**Fig. 5c**) and is responsible for the intra-speed settling rate variation (**Fig. 5d**).

## Discussion

Our results demonstrate that physical larval characteristics can produce markedly different settlement outcomes under identical flow and substrate conditions, with settlement success varying up to an order of magnitude across trait combinations. Crucially, neither swimming speed nor body shape is consistently the primary determinant of settlement. Instead, substrate topography-flow interactions dictate whether settlement hydrodynamics are governed by swimming speed or shape. Near-surface flow acts as a passive hydrodynamic trait filter, regulating which larvae can cross high-velocity transport barriers, remain retained near the substrate, and access sheltered settlement microrefugia.

This trait-filtering mechanism operates differently across flow regimes (**Fig. 1b**). In skimming flow, high velocities above the ridge tops isolate bulk flow from recirculating flow within gaps, forming a transport barrier that restricts entry into sheltered regions. Faster larvae cross this barrier more often, increasing settlement both inside protected gaps and on exposed ridge tops. Reattached flow instead allows fluid to enter the wider gaps, but near-surface flow speeds exceed 5× the fastest larval swimming speed, creating a settlement barrier across the central gap floor. Faster larvae overcome this barrier by exploiting the brief windows when local flow drops below their settlement threshold. In both regimes, swimming speed is therefore the primary determinant of settlement success, though the barrier it must overcome differs.

In contrast, detached flow alters the trait filter by shifting primary control from swimming speed to body shape. Here, periodic vortex ejection connects the bulk flow to the gap interior. Higher aspect ratio, or elongated, larvae couple more strongly to local strain fields, aligning with flow streamlines and enhancing net transport into sheltered ridge gaps (**Fig. S6** and **Video S1**). Swimming speed, however, governs a trade-off within the gaps: slower larvae are more effectively trapped by recirculating flow therefore increasing near-surface accumulation, while faster larvae settle at higher fractional rates once inside. These opposing effects partially offset, leading to a maximum settlement rate for elongated larvae ($\alpha \geq 0.5$) at intermediate swimming speeds ($u_l = 2$ mm $s^{-1}$) (**Fig. 5d**). Swimming speeds and body shapes near this simulated maximum have been reported for coral larvae, including both traits in the Caribbean reef-building coral *Colpophyllia natans* (**Table S2**), motivating further consideration of how substrate topography might favor larvae with particular trait combinations.

These findings demonstrate that substrate topography, hydrodynamics, and organismal traits cannot be evaluated as independent controls on recruitment. Indeed, this suggests that substrate topography could be used to alter physical access to settlement surfaces for larvae with particular characteristics under known flow conditions. Such a principle may be useful in restoration, where substrates are often intended to promote establishment by target taxa, including keystone, fast-growing, or pioneer species that contribute to ecosystem recovery [55–57]. It could also inform strategies intended to reduce access by unwanted fouling organisms [14,58–60].

This same physical understanding could also help explain settlement patterns and predict settlement success in natural systems more broadly. However, applying this framework will require local characterization of substrate roughness and hydrodynamics, together with biological validation of whether target species settle as predicted, since behavioral and habitat cues fall outside of the physical model. Field-resolvable flow metrics that identify flow separation, recirculation, and near-surface transport intensity, including the Q-criterion or vertical velocity variance [24,41], may help link the mechanisms identified here to larger-scale models, site selection, and substrate design. Similarly, these results may find use in updating large-scale larval dispersal and population connectivity models, which should treat biological traits and hydrodynamics interdependently [3,61], and which often do not consider fine-scale aspects of larval settlement [62,63] or have the data required for parameterizing larval traits [61].

Our mechanistic model isolates the physical controls on settlement by introducing deliberate biological and environmental simplifications. The simulations are two-dimensional and use idealized substrate topographies and larval shapes. Larvae swim at constant speed, are neutrally buoyant, and do not alter their motion in response to hydrodynamic, chemical, or biological cues. Settlement is represented by a velocity-based threshold and therefore does not account for substrate properties, attachment mechanics, or post-attachment survival. The flow simulations consider a single wave-driven oscillatory-flow condition for a relatively short timescale and do not include unidirectional currents or longer-range transport to the substrate. These assumptions and simplifications—made within an ecologically-relevant and physically reasonable range of conditions [41]—allowed us to examine swimming speed and body shape independently over a large parameter space. This was necessary to isolate the short-range hydrodynamic mechanisms that govern physical access to settlement surfaces while limiting species-specific behavioral assumptions. Future extensions of this framework that incorporate three-dimensional non-idealized topographies, turbulent flow fluctuations, active behavioral responses, substrate attachment dynamics, and settler survival will be necessary to determine when physical access translates into realized recruitment.

In summary, near-surface flow regimes dictate whether larval settlement hydrodynamics are governed by swimming speed or body shape. Faster swimming enables larvae to breach rapid-flow transport and settlement barriers, whereas elongated morphology enhances alignment and transport within recirculating vortical flows. By demonstrating how millimeter-scale hydrodynamics filter organismal traits, these results provide a physical foundation for understanding, and potentially guiding, larval recruitment in natural benthic ecosystems.

**Funding**: This work was supported by the National Science Foundation (IOS 1848671).

**Acknowledgments**:
The authors thank their previous team of collaborators, including: Amy Wagoner Johnson, Joaquin Yus, Kristen Marhaver, Rosa Espinosa Marzal, Zachary Quinlan and Linda Wegley Kelly for helpful discussions. We also thank Valérie Chamberland and Robbert-Jan Geertsma for larval sizing information.

## References

1. Ritson-Williams R, Arnold S, Fogarty N, Steneck RS, Vermeij M, Paul VJ. 2009 New perspectives on ecological mechanisms affecting coral recruitment on reefs. *Smithson. Contrib. Mar. Sci.* , 437–457. (doi:10.5479/si.01960768.38.437)

2. Cowen RK, Sponaugle S. 2009 Larval Dispersal and Marine Population Connectivity. *Ann. Rev. Mar. Sci.* **1**, 443–466. (doi:10.1146/annurev.marine.010908.163757)

3. Álvarez-Noriega M, Burgess SC, Byers JE, Pringle JM, Wares JP, Marshall DJ. 2020 Global biogeography of marine dispersal potential. *Nat. Ecol. Evol.* **4**, 1196–1203. (doi:10.1038/s41559-020-1238-y)

4. Yahagi T, Tanaka K, Higuchi T, Shirai K, Takahata N, Sano Y, Kojima S, Kano Y. 2026 Gastropod shells record larval migration from deep-sea hydrothermal vents to the euphotic zone. *Sci. Adv.* **12**, 7045. (doi:10.1126/sciadv.adx7045)

5. Koehl MAR, Hadfield MG. 2010 Hydrodynamics of Larval Settlement from a Larva's Point of View. *Integr. Comp. Biol.* **50**, 539–551. (doi:10.1093/icb/icq101)

6. Tebben J *et al.* 2015 Chemical mediation of coral larval settlement by crustose coralline algae. *Sci. Rep.* **5**, 10803. (doi:10.1038/srep10803)

7. Gómez-Lemos LA, Doropoulos C, Bayraktarov E, Diaz-Pulido G. 2018 Coralline algal metabolites induce settlement and mediate the inductive effect of epiphytic microbes on coral larvae. *Sci. Rep.* **8**, 17557. (doi:10.1038/s41598-018-35206-9)

8. Levenstein MA *et al.* 2022 Composite Substrates Reveal Inorganic Material Cues for Coral Larval Settlement. *ACS Sustain. Chem. Eng.* **10**, 3960–3971. (doi:10.1021/acssuschemeng.1c08313)

9. Quinlan ZA *et al.* 2023 Coral larval settlement induction using tissue-associated and exuded coralline algae metabolites and the identification of putative chemical cues. *Proceedings of the Royal Society B: Biological Sciences* **290**. (doi:10.1098/rspb.2023.1476)

10. Yus J *et al.* 2024 Composite substrates for coral larval settlement and reef restoration based on natural hydraulic lime and inorganic strontium and magnesium compounds. *Ecol. Eng.* **202**, 107236. (doi:10.1016/j.ecoleng.2024.107236)

11. Ruszczyk M *et al.* 2026 Alkalinity-enhanced artificial substrates modulate local pH and increase survivorship of early-stage coral recruits. *Commun. Earth Environ.* **7**, 311. (doi:10.1038/s43247-026-03414-1)

12. Ramon-Mateu J, Ferraioli A, Teixidó N, Domart-Coulon I, Houliston E, Copley RR. 2025 Aboral cell types of *Clytia* and coral larvae have shared features and link taurine to the regulation of settlement. *Sci. Adv.* **11**, 1159. (doi:10.1126/sciadv.adv1159)

13. Callow ME, Jennings AR, Brennan AB, Seegert CE, Gibson A, Wilson L, Feinberg A, Baney R, Callow JA. 2002 Microtopographic Cues for Settlement of Zoospores of the Green Fouling Alga Enteromorpha. *Biofouling* **18**, 229–236. (doi:10.1080/08927010290014908)

14. Scardino AJ, Guenther J, de Nys R. 2008 Attachment point theory revisited: the fouling response to a microtextured matrix. *Biofouling* **24**, 45–53. (doi:10.1080/08927010701784391)

15. Myan FWY, Walker J, Paramor O. 2013 The interaction of marine fouling organisms with topography of varied scale and geometry: a review. *Biointerphases* **8**, 30-. (doi:10.1186/1559-4106-8-30)

16. Carl C, Poole AJ, Sexton BA, Glenn FL, Vucko MJ, Williams MR, Whalan S, de Nys R. 2012 Enhancing the settlement and attachment strength of pediveligers of *Mytilus galloprovincialis* bychanging surface wettability and microtopography. *Biofouling* **28**, 175–186. (doi:10.1080/08927014.2012.662676)

17. Whalan S, Abdul Wahab MA, Sprungala S, Poole AJ, de Nys R. 2015 Larval Settlement: The Role of Surface Topography for Sessile Coral Reef Invertebrates. *PLoS One* **10**, e0117675. (doi:10.1371/journal.pone.0117675)

18. Erramilli S, Genzer J. 2019 Influence of surface topography attributes on settlement and adhesion of natural and synthetic species. *Soft Matter* **15**, 4045–4067. (doi:10.1039/C9SM00527G)

19. Hata T, Madin JS, Cumbo VR, Denny M, Figueiredo J, Harii S, Thomas CJ, Baird AH. 2017 Coral larvae are poor swimmers and require fine-scale reef structure to settle. *Sci. Rep.* **7**, 2249. (doi:10.1038/s41598-017-02402-y)

20. Davis KA, Pawlak G, Monismith SG. 2021 Turbulence and Coral Reefs. *Ann. Rev. Mar. Sci.* **13**, 343–373. (doi:10.1146/annurev-marine-042120-071823)

21. Koehl MAR, Cooper T. 2015 Swimming in an Unsteady World. *Integr. Comp. Biol.* **55**, 683–697. (doi:10.1093/icb/icv092)

22. Whitman E, Reidenbach M. 2012 Benthic flow environments affect recruitment of Crassostrea virginica larvae to an intertidal oyster reef. *Mar. Ecol. Prog. Ser.* **463**, 177–191. (doi:10.3354/meps09882)

23. Larsson AI, Granhag LM, Jonsson PR. 2016 Instantaneous Flow Structures and Opportunities for Larval Settlement: Barnacle Larvae Swim to Settle. *PLoS One* **11**, e0158957. (doi:10.1371/journal.pone.0158957)

24. Levenstein MA *et al.* 2022 Millimeter-scale topography facilitates coral larval settlement in wave-driven oscillatory flow. *PLoS One* **17**, e0274088. (doi:10.1371/journal.pone.0274088)

25. Lanham BS, Pomeroy AWM, Swearer SE, Marusic I, Javaherchian J, Morris RL. 2025 Substrate-mediated alterations to hydrodynamic conditions enhances shellfish larval settlement: Implications for artificial reef restoration. *Ecol. Eng.* **212**, 107474. (doi:10.1016/j.ecoleng.2024.107474)

26. Gaylord B, Hodin J, Ferner MC. 2013 Turbulent shear spurs settlement in larval sea urchins. *Proceedings of the National Academy of Sciences* **110**, 6901–6906. (doi:10.1073/pnas.1220680110)

27. Fuchs HL, Gerbi GP, Hunter EJ, Christman AJ, Diez FJ. 2015 Hydrodynamic sensing and behavior by oyster larvae in turbulence and waves. *Journal of Experimental Biology* **218**, 1419–1432. (doi:10.1242/jeb.118562)

28. Michalec F-G, Fouxon I, Souissi S, Holzner M. 2017 Zooplankton can actively adjust their motility to turbulent flow. *Proceedings of the National Academy of Sciences* **114**, E11199–E11207. (doi:10.1073/pnas.1708888114)

29. Monthiller R, Loisy A, Koehl MAR, Favier B, Eloy C. 2022 Surfing on Turbulence: A Strategy for Planktonic Navigation. *Phys. Rev. Lett.* **129**, 064502. (doi:10.1103/PhysRevLett.129.064502)

30. Deshpande K *et al.* In press. Direct observation and quantitative characterization of chemotactic behaviors in Caribbean coral larvae exposed to organic and inorganic settlement cues. (doi:10.1038/s41598-025-93194-z)

31. Chamberland VF, Snowden S, Marhaver KL, Petersen D, Vermeij MJA. 2017 The reproductive biology and early life ecology of a common Caribbean brain coral, Diploria labyrinthiformis (Scleractinia: Faviinae). *Coral Reefs* **36**, 83–94. (doi:10.1007/s00338-016-1504-2)

32. Weinberg S, Weinberg F. 1979 The Life Cycle of a Gorgonian: Eunicella Singularis (Esper, 1794). *Bijdragen tot de Dierkunde* **48**, 127–140. (doi:10.1163/26660644-04802003)

33. Berman SA, Mitchell KA. 2020 Trapping of swimmers in a vortex lattice. *Chaos: An Interdisciplinary Journal of Nonlinear Science* **30**. (doi:10.1063/5.0005542)

34. Berman SA, Buggeln J, Brantley DA, Mitchell KA, Solomon TH. 2021 Transport barriers to self-propelled particles in fluid flows. *Phys. Rev. Fluids* **6**, L012501. (doi:10.1103/PhysRevFluids.6.L012501)

35. Si X, Fang L. 2022 Preferential transport of swimmers in heterogeneous two-dimensional turbulent flow. *Phys. Rev. Fluids* **7**, 094501. (doi:10.1103/PhysRevFluids.7.094501)

36. Zhang Z, Qiu J, Zhao L. 2022 Micro-swimmers in vertical turbulent channel flows. *International Journal of Multiphase Flow* **151**, 104035. (doi:10.1016/j.ijmultiphaseflow.2022.104035)

37. Qin B, Arratia PE. 2022 Confinement, chaotic transport, and trapping of active swimmers in time-periodic flows. *Sci. Adv.* **8**. (doi:10.1126/sciadv.add6196)

38. Tanasijević I, Lauga E. 2022 Microswimmers in vortices: dynamics and trapping. *Soft Matter* **18**, 8931–8944. (doi:10.1039/D2SM00907B)

39. Rusconi R, Guasto JS, Stocker R. 2014 Bacterial transport suppressed by fluid shear. *Nat. Phys.* **10**, 212–217. (doi:10.1038/nphys2883)

40. Lovecchio S, Climent E, Stocker R, Durham WM. 2019 Chain formation can enhance the vertical migration of phytoplankton through turbulence. *Sci. Adv.* **5**. (doi:10.1126/sciadv.aaw7879)

41. Gysbers D, Levenstein MA, Juarez G. 2025 Effect of substrate topography on benthic boundary layer flow: Implications for marine larval transport and settlement. *Phys. Rev. Fluids* **10**, 064501. (doi:10.1103/4dp1-36d8)

42. Antonio-Martínez F, Henaut Y, Vega-Zepeda A, Cerón-Flores AI, Raigoza-Figueras R, Cetz-Navarro NP, Espinoza-Avalos J. 2020 Leachate effects of pelagic Sargassum spp. on larval swimming behavior of the coral Acropora palmata. *Sci. Rep.* **10**, 3910. (doi:10.1038/s41598-020-60864-z)

43. Heran T, Laudien J, Waller RG, Häussermann V, Försterra G, González HE, Richter C. 2023 Life cycle of the cold-water coral Caryophyllia huinayensis. *Sci. Rep.* **13**, 2593. (doi:10.1038/s41598-023-29620-x)

44. Villanueva RD, Baria MVB, dela Cruz DW, Dizon RM. 2011 Diel timing of planulation and larval settlement in the coral Isopora cuneata (Scleractinia: Acroporidae). *Hydrobiologia* **673**, 273–279. (doi:10.1007/s10750-011-0771-8)

45. Poon RN, Westwood TA, Laeverenz-Schlogelhofer H, Brodrick E, Craggs J, Keaveny EE, Jékely G, Wan KY. 2023 Ciliary propulsion and metachronal coordination in reef coral larvae. *Phys. Rev. Res.* **5**, L042037. (doi:10.1103/PhysRevResearch.5.L042037)

46. Ferrari S, Cribari-Neto F. 2004 Beta Regression for Modelling Rates and Proportions. *J. Appl. Stat.* **31**, 799–815. (doi:10.1080/0266476042000214501)

47. Hastie T, Tibshirani R, Friedman J. 2009 *The Elements of Statistical Learning*. New York, NY: Springer New York. (doi:10.1007/978-0-387-84858-7)

48. Josef Perktold, Skipper Seabold, Jonathan Taylor, statsmodels-developers. In press. statsmodels.othermod.betareg.BetaModel documentation. See https://www.statsmodels.org/devel/generated/statsmodels.othermod.betareg.BetaModel.html (accessed on 19 August 2026).

49. Roberts DR *et al.* 2017 Cross-validation strategies for data with temporal, spatial, hierarchical, or phylogenetic structure. *Ecography* **40**, 913–929. (doi:10.1111/ecog.02881)

50. Seabold S, Perktold J. 2010 Statsmodels: Econometric and Statistical Modeling with Python. In *SciPy 2010*, pp. 92–96. SciPy. (doi:10.25080/Majora-92bf1922-011)

51. McKinney W. 2010 Data Structures for Statistical Computing in Python. In *SciPy 2010*, pp. 56–61. SciPy. (doi:10.25080/Majora-92bf1922-00a)

52. Harris CR *et al.* 2020 Array programming with NumPy. *Nature* **585**, 357–362. (doi:10.1038/s41586-020-2649-2)

53. Virtanen P *et al.* 2020 SciPy 1.0: fundamental algorithms for scientific computing in Python. *Nat. Methods* **17**, 261–272. (doi:10.1038/s41592-019-0686-2)

54. Pedregosa F *et al.* 2011 Scikit-learn: Machine Learning in Python. *The Journal of Machine Learning Research* (doi:10.5555/1953048.2078195)

55. Madin JS *et al.* 2023 Selecting coral species for reef restoration. *Journal of Applied Ecology* **60**, 1537–1544. (doi:10.1111/1365-2664.14447)

56. Hale SL, Koprowski JL. 2018 Ecosystem-level effects of keystone species reintroduction: a literature review. *Restor. Ecol.* **26**, 439–445. (doi:10.1111/rec.12684)

57. Boström-Einarsson L *et al.* 2020 Coral restoration – A systematic review of current methods, successes, failures and future directions. *PLoS One* **15**, e0226631. (doi:10.1371/journal.pone.0226631)

58. Vucko MJ, Poole AJ, Carl C, Sexton BA, Glenn FL, Whalan S, de Nys R. 2014 Using textured PDMS to prevent settlement and enhance release of marine fouling organisms. *Biofouling* **30**, 1–16. (doi:10.1080/08927014.2013.836507)

59. Berntsson KM, Jonsson PR, Lejhall M, Gatenholm P. 2000 Analysis of behavioural rejection of micro-textured surfaces and implications for recruitment by the barnacle Balanus improvisus. *J. Exp. Mar. Biol. Ecol.* **251**, 59–83. (doi:10.1016/S0022-0981(00)00210-0)

60. Efimenko K, Finlay J, Callow ME, Callow JA, Genzer J. 2009 Development and Testing of Hierarchically Wrinkled Coatings for Marine Antifouling. *ACS Appl. Mater. Interfaces* **1**, 1031–1040. (doi:10.1021/am9000562)

61. Ani C, Grimaldi C, Iwanaga T, Treml E, Robson B. 2026 Coral connectivity: How biophysical modelling choices shape connectivity inference. (doi:10.31223/X5022V)

62. Hock K, Doropoulos C, Gorton R, Condie SA, Mumby PJ. 2019 Split spawning increases robustness of coral larval supply and inter-reef connectivity. *Nat. Commun.* **10**, 3463. (doi:10.1038/s41467-019-11367-7)

63. Bozec Y-M *et al.* 2025 A rapidly closing window for coral persistence under global warming. *Nat. Commun.* **16**, 9704. (doi:10.1038/s41467-025-65015-4)

**Figures:**

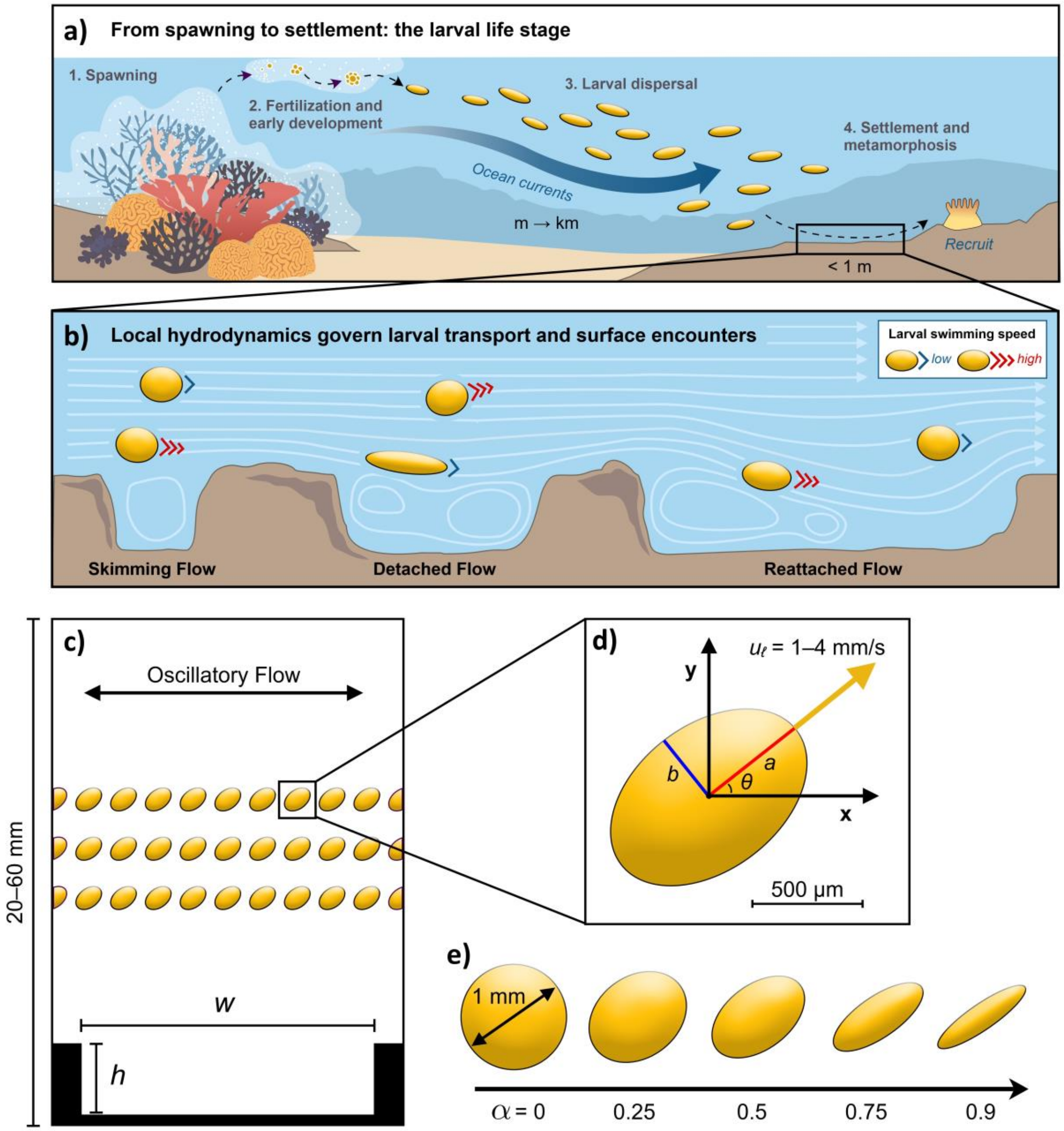

**Figure 1. Overview of larval dispersal and settlement hydrodynamics and implementation of the mechanistic model. (a)** Early life history of marine invertebrate larvae from spawning to settlement. **(b)** Illustration of near-surface larval transport hydrodynamics including the three topography-induced flow regimes: skimming, detached, and reattached, shown during the max flow phase of the oscillatory period. Larvae are not to scale. **(c)** The simulated flow region where the ridge height (*h*) and spacing (*w*) were varied to investigate the effects of substrate topography.

Initial larval positions are illustrated, and each larva is simulated independently. The domain height was restricted to 20 mm for computational efficiency, but some simulations were conducted with heights of up to 60 mm to account for vortices ejected high into the bulk flow. **(d)** A model larva used in settlement simulations with characteristic parameters indicated. **(e)** Larval body shapes with varying parameter α.

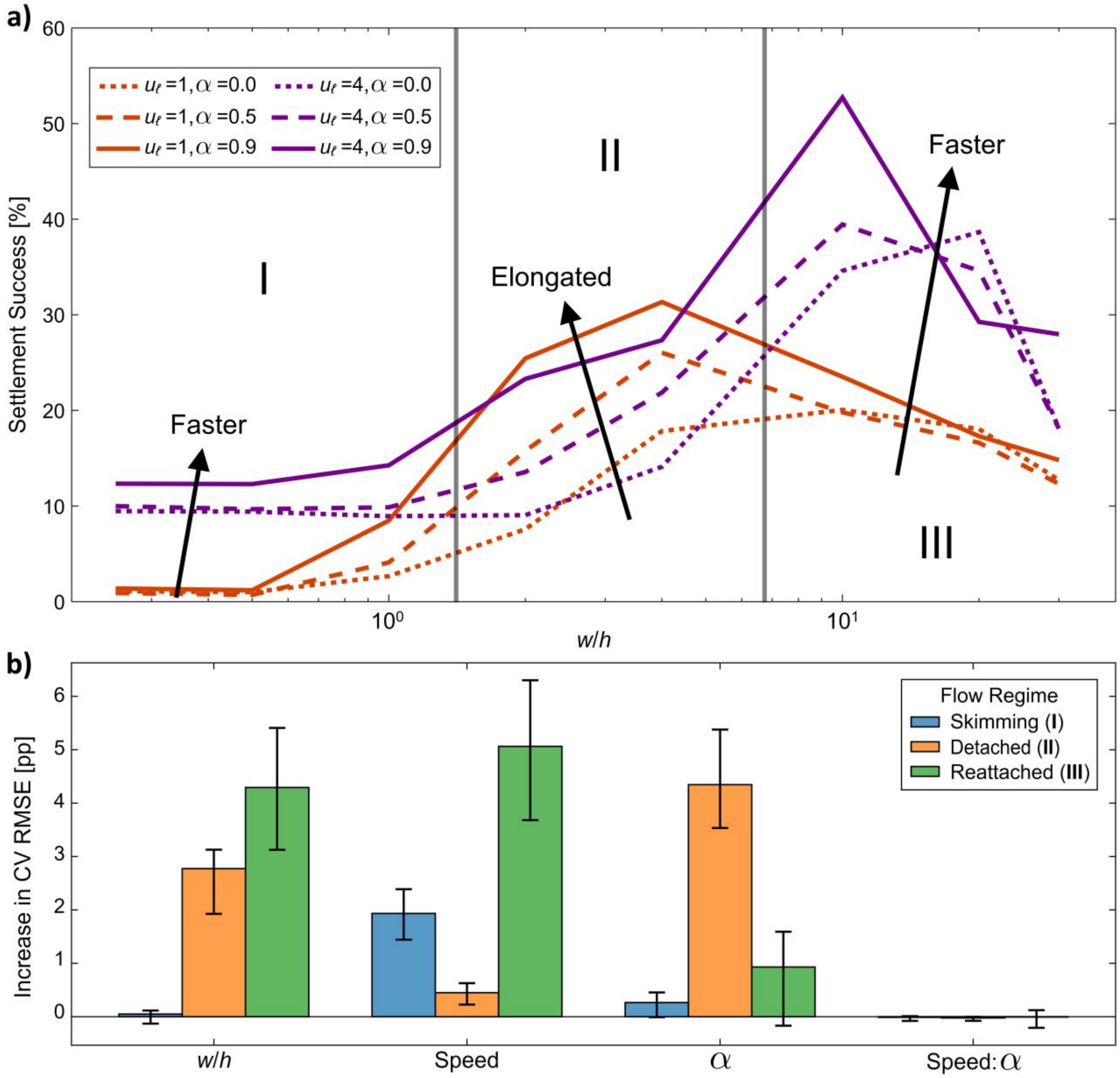

**Figure 2**. **Larval settlement in different flow regimes. (a)** Settlement percentages for a range of larval parameters ($u_l$, α) and *w*/*h* values. Grey lines separate substrates by the observed flow regimes: **(I)** skimming flow, **(II)** detached flow, and **(III)** reattached flow. In each flow regime, the larval characteristic that most influences settlement success is indicated. **(b)** Predictive importance of *w*/*h*, speed, shape parameter ($\alpha$), and the speed × $\alpha$ interaction within each flow regime, quantified as the median increase in grouped CV RMSE in percentage points (pp) after hierarchical removal of the corresponding parameter from the beta regression model. Positive values indicate that inclusion of the parameter improves out-of-sample prediction, while error bars show the 25-75% interquartile range (IQR) across paired repeated CV folds.

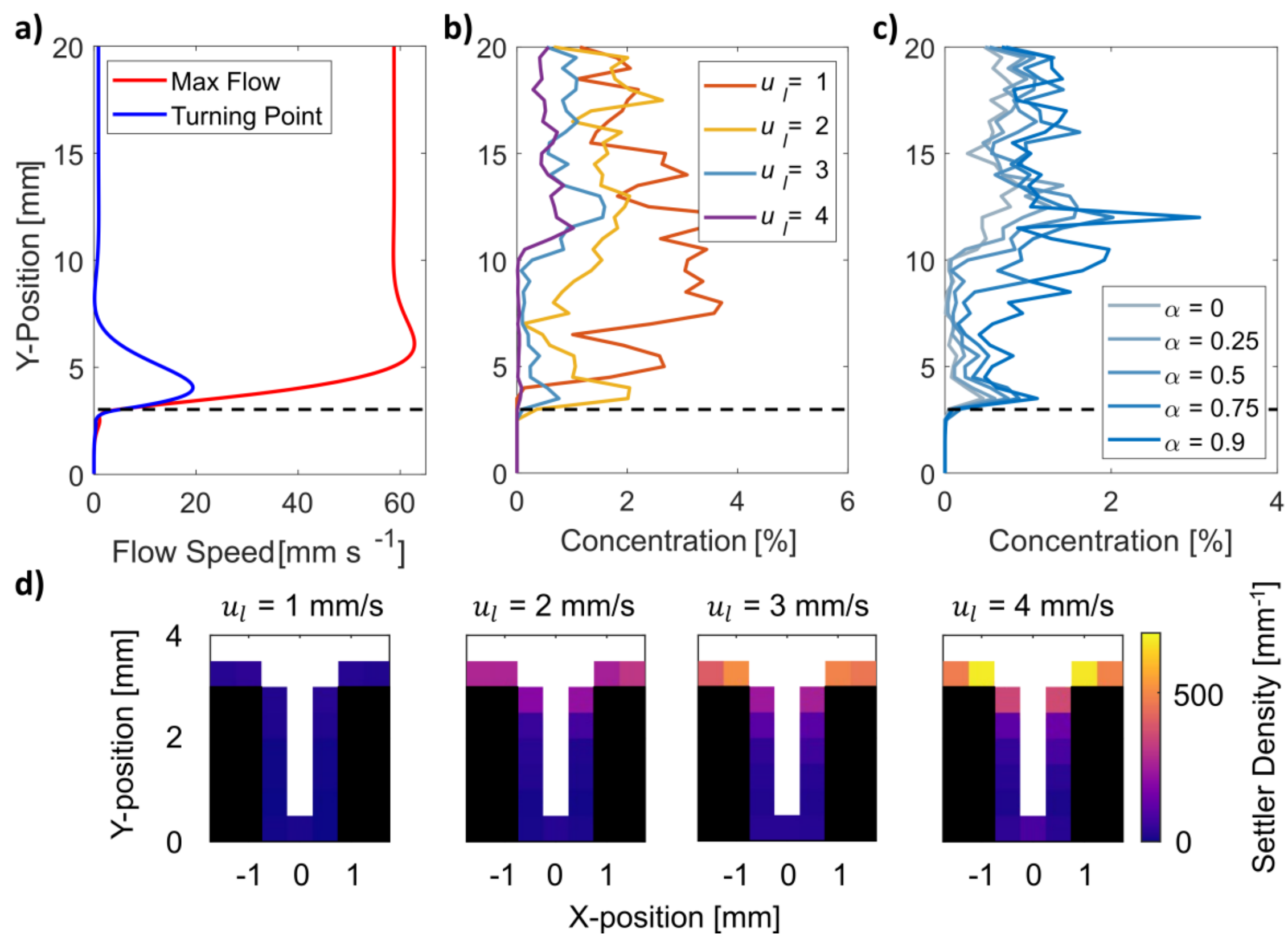

**Figure 3**. **Flow speeds, larval concentration profiles, and settlement locations for skimming flow. (a)** The simulated flow speed profile during the max flow and turning point of the oscillatory period for the substrate with *w*/*h* = 0.5 in the skimming flow regime. **(b)** Larval concentration profiles above the *w*/*h* = 0.5 substrate for larvae with α = 0.5 and four different swimming speeds after three periods (16.5 s) of oscillatory flow. **(c)** Larval concentration profiles above the *w*/*h* = 0.5 substrate for larvae with $u_l$ = 3 mm $s^{-1}$ and five different α after three periods (16.5 s) of oscillatory flow. **(d)** Settlement heatmaps for the *w*/*h* = 0.5 substrate for larvae with α = 0.5.

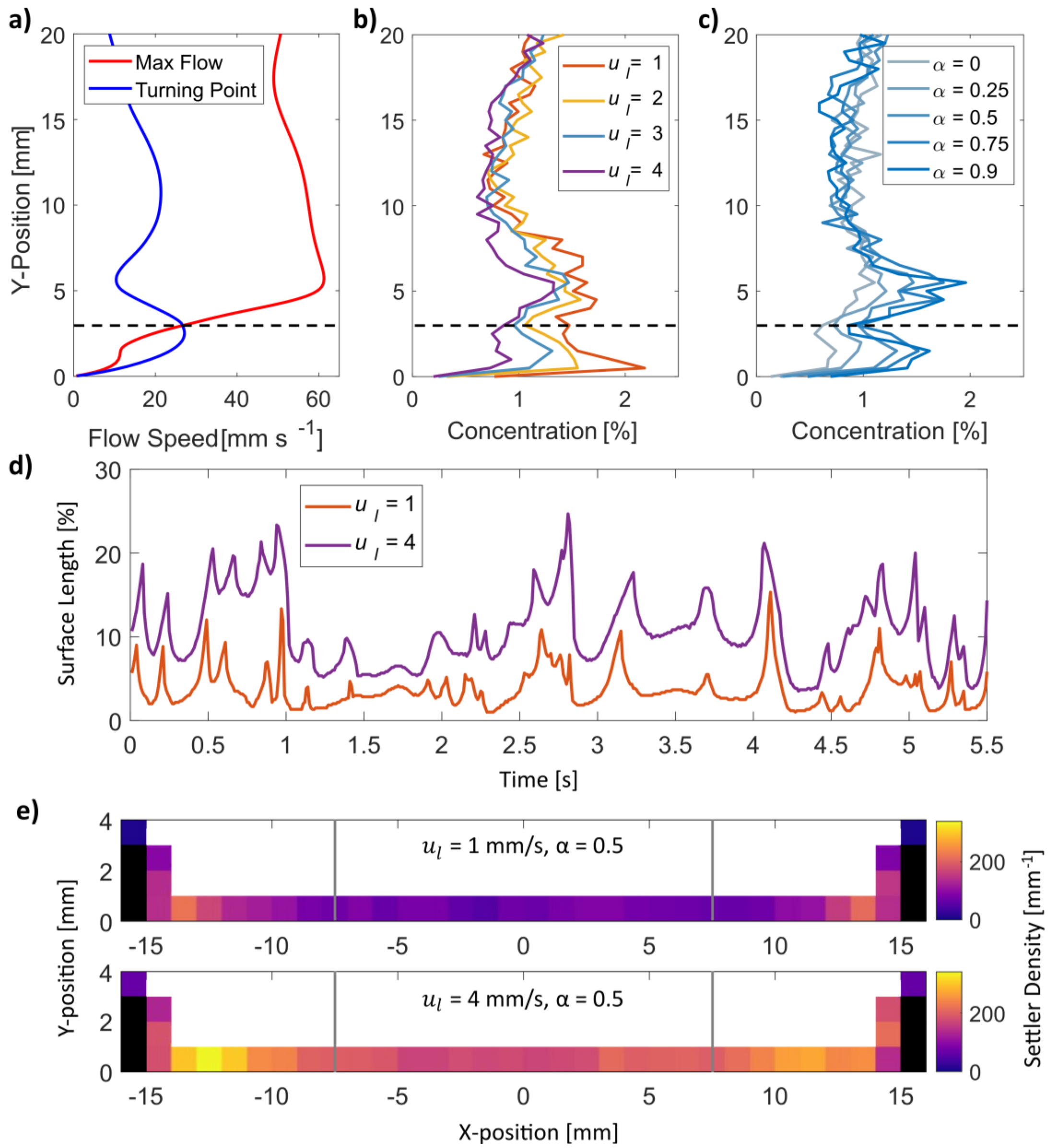

**Figure 4**. **Flow speeds, larval concentration profiles, and settlement locations for reattached flow. (a)** The simulated flow speed profile during the max flow and turning point of the oscillatory period for the substrate with *w/h* = 10 in the reattached flow regime. **(b)** Larval concentration profiles above the *w/h* = 10 substrate for larvae with α = 0.5 and four different swimming speeds after three periods (16.5 s) of oscillatory flow. **(c)** Larval concentration profiles above the *w/h* = 10 substrate for larvae with $u_l$ = 3 mm $s^{-1}$ and five different α after three periods (16.5 s) of oscillatory flow. **(d)** Percentage of the surface length of the *w/h* = 10 substrate that is available for settlement for an example period. The surface is considered available for settlement if it has a flow speed of less than 1.33× the larval swimming speed at half a larval body length from the surface.

**(e)** Settlement heatmaps for the *w*/*h* = 10 substrate. The center 50% of the surface and 25% at each corner are separated by grey lines.

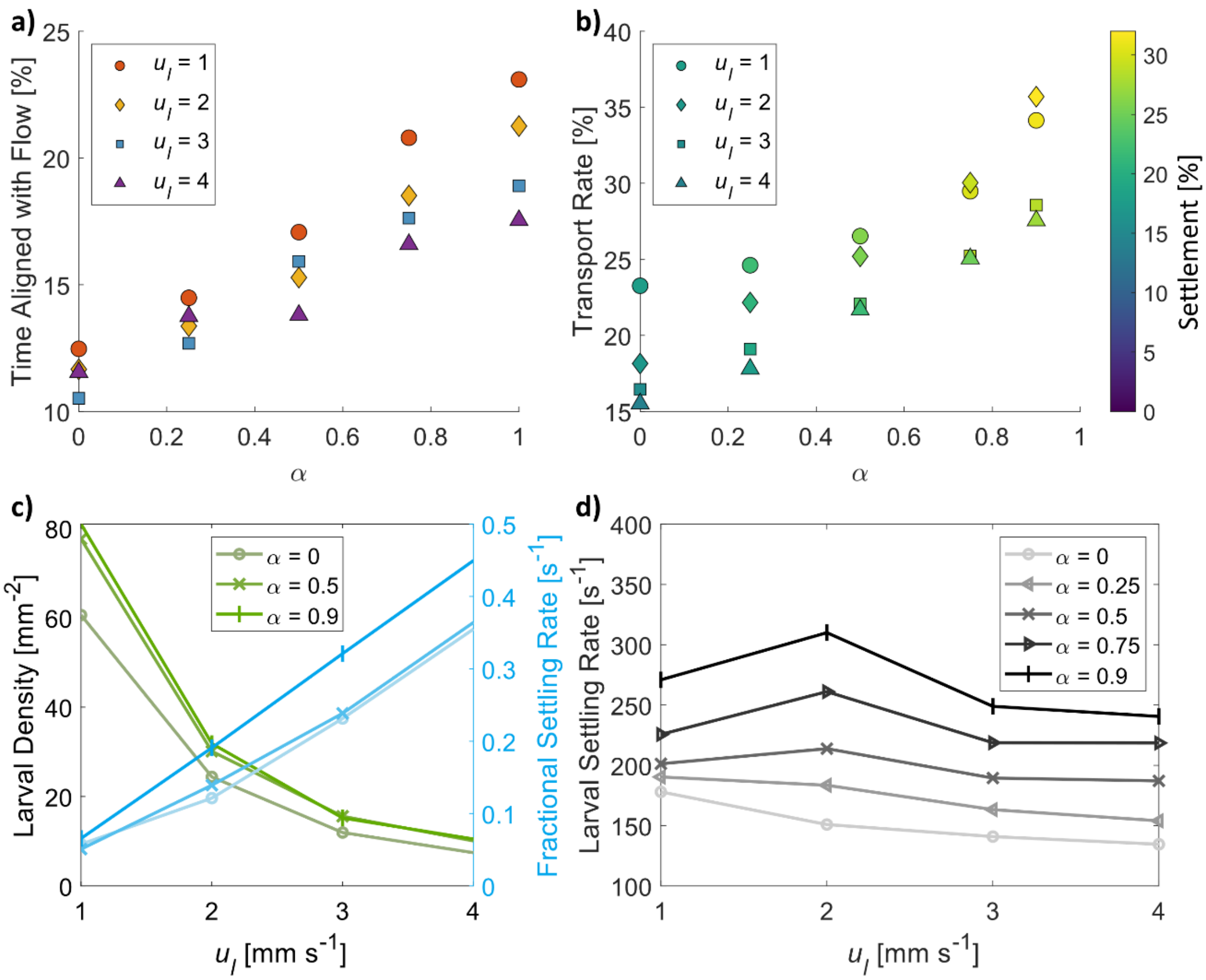

**Figure 5**. **Larval alignment, transport rate, and settling rates for the detached flow regime.** **(a)** Average time larvae are aligned with the local flow direction within 5 mm of the surface as a percentage of their time in that region for a substrate with $w/h = 4$. Larvae are considered aligned when their director points within 0.2 radians of the flow direction. **(b)** Larval transport rate into the ridge gaps as a percentage of the rate of larvae entering the simulation. **(c)** Larval density within the ridge gaps and the larval settling rate as a fraction of the number of larvae within the gaps. **(d)** Overall larval settling rate for larvae with five different α values at four swimming speeds.